\documentclass[aps,prl,twocolumn,superscriptaddress,showpacs]{revtex4}
\usepackage{amsmath}
\usepackage{amssymb}
\usepackage{colordvi}
\usepackage{mathrsfs}
\usepackage{bm}
\usepackage{verbatim}
\usepackage{dcolumn}
\usepackage{bm}
\usepackage{epsfig}
\usepackage{subfigure}
\begin{document}
\title{Anderson Localization from Berry-Curvature Interchange in Quantum Anomalous Hall System}
\author{Zhenhua Qiao}
\affiliation{ICQD, Hefei National Laboratory for Physical Sciences at Microscale, and Synergetic Innovation Center of Quantum Information and Quantum Physics, University of Science and Technology of China, Hefei, Anhui 230026, China.}
\affiliation{CAS Key Laboratory of Strongly-Coupled Quantum Matter Physics, and Department of Physics, University of Science and Technology of China, Hefei, Anhui 230026, China.}
\affiliation{Department of Physics, The University of Texas at Austin, Austin, Texas 78712, USA}
\author{Ke Wang}
\affiliation{ICQD, Hefei National Laboratory for Physical Sciences at Microscale, and Synergetic Innovation Center of Quantum Information and Quantum Physics, University of Science and Technology of China, Hefei, Anhui 230026, China.}
\affiliation{CAS Key Laboratory of Strongly-Coupled Quantum Matter Physics, and Department of Physics, University of Science and Technology of China, Hefei, Anhui 230026, China.}
\author{Lei Zhang}
\affiliation{Department of Physics and the Center of Theoretical and Computational Physics, The University of Hong Kong, Pokfulam Road, Hong Kong, China}
\author{Yulei Han}
\affiliation{ICQD, Hefei National Laboratory for Physical Sciences at Microscale, and Synergetic Innovation Center of Quantum Information and Quantum Physics, University of Science and Technology of China, Hefei, Anhui 230026, China.}
\affiliation{CAS Key Laboratory of Strongly-Coupled Quantum Matter Physics, and Department of Physics, University of Science and Technology of China, Hefei, Anhui 230026, China.}
\author{Xinzhou Deng}
\affiliation{ICQD, Hefei National Laboratory for Physical Sciences at Microscale, and Synergetic Innovation Center of Quantum Information and Quantum Physics, University of Science and Technology of China, Hefei, Anhui 230026, China.}
\affiliation{CAS Key Laboratory of Strongly-Coupled Quantum Matter Physics, and Department of Physics, University of Science and Technology of China, Hefei, Anhui 230026, China.}
\author{Hua Jiang}
\affiliation{College of Physics, Soochou University, Suzhou, Jiangsu 215006, China}
\affiliation{Department of Physics, The University of Texas at Austin, Austin, Texas 78712, USA}
\author{Shengyuan A. Yang}
\affiliation{Research Laboratory for Quantum Materials, Singapore University of Technology and Design, Singapore 487372, Singapore}
\affiliation{Department of Physics, The University of Texas at Austin, Austin, Texas 78712, USA}
\author{Jian Wang}
\affiliation{Department of Physics and the Center of Theoretical and Computational Physics, The University of Hong Kong, Pokfulam Road, Hong Kong, China}
\author{Qian Niu}
\affiliation{Department of Physics, The University of Texas at Austin, Austin, Texas 78712, USA}
\date{\today}
\begin{abstract}
  We theoretically investigate the localization mechanism of the quantum anomalous Hall effect (QAHE) in the presence of spin-flip disorders. We show that the QAHE keeps quantized at weak disorders, then enters a Berry-curvature mediated metallic phase at moderate disorders, and finally goes into the Anderson insulating phase at strong disorders. From the phase diagram, we find that at the charge neutrality point although the QAHE is most robust against disorders, the corresponding metallic phase is much easier to be localized into the Anderson insulating phase due to the \textit{interchange} of Berry curvatures carried respectively by the conduction and valence bands. At the end, we provide a phenomenological picture related to the topological charges to better understand the underlying physical origin of the QAHE Anderson localization.
\end{abstract}
\pacs{73.43.Cd, 
          73.43.-f,   
          73.23.-b,  
          71.30.+h. 
         }
\maketitle
\textit{Introduction.---} The Anderson localization~\cite{Anderson} is one of the most striking transport phenomena in condensed matter physics. It describes the absence of diffusion of waves due to the severe interference from strong disorders. Based on the scaling theory of the localization length, it is known that two-dimensional electrons can be immediately driven into the Anderson insulating phase even in the presence of extremely weak disorders~\cite{ScalingTheory}. However, if either the time-reversal symmetry is broken by the magnetic field or the spin-rotational symmetry is broken by the spin-orbit couplings, a metal-insulator phase transition occurs~\cite{RMP-Beenakker,UnitaryClass1,UnitaryClass2,UnitaryClass3,UnitaryClass4}, indicating the emergence of a metallic phase at weak disorders.

When the applied magnetic field is strong enough, the resulting Landau-level quantization gives rise to the formation of the conventional quantum Hall effect~\cite{QHE1,QHE2}, manifesting itself as vanishing longitudinal conductance but quantized Hall conductance. In the presence of disorders, there were several different localization mechanisms proposed for the quantum Hall effect, \textit{e.g.}, a levitation theory where extended levels float up to infinity at weak magnetic field limit was used to show that the phase transition can only occur in nearest-neighbor quantum Hall plateaus, indicating that a high QHE state can not directly transit into an insulator~\cite{GlobalPhaseSCZ}, while Sheng et al. suggested that the integer quantum Hall plateaus are destroyed in a one-by-one order from high to low energies without floating up in energy~\cite{DisappearIQHEDNS}. When interband mixing effect of opposite chiralities are considered, metallic phase may exist between adjacent quantum Hall plateaus or between quantum Hall phase and an Anderson insulator~\cite{IQHEGXiong}. Recently, the successful exfoliation of  monolayer graphene~\cite{graphene} and realization of $\mathbb{Z}_2$ topological insulators~\cite{TI1,TI2} (both harbour linear-Dirac dispersions) inspire a broad exploration of the quantum anomalous Hall effect (QAHE) in related materials~\cite{QAHE_Review,QAHE1,QAHE2,QAHE3,QAHE4,QAHE5,QAHE6,QAHE7,QAHE8,QAHE9,QAHE10,QAHE11,QAHE12,QAHE13} and finally lead to the first realization of QAHE in magnetically doped topological insulators without applying an external magnetic field~\cite{QAHEExp1,QAHEExp2,QAHEExp3,QAHEExp4}, where the QAHE exhibits the same transport properties as those in the conventional quantum Hall effect. The formation of QAHE usually originates from the synergetic interaction between the spin-orbit coupling and intrinsic magnetization. Therefore, a natural and fundamental question arises considering that both the time-reversal and spin-rotational symmetries are broken: how will the QAHE phase be localized in the presence of strong disorders?

In this Letter, we first investigate the transport properties of the chiral edge states of the QAHE in the presence of both nonmagnetic and spin-flip disorders. In the nonmagnetic case, for Fermi-levels lying inside the band gap, the conductance is quantized at weak disorders, gradually decreases at moderate disorders, and finally vanishes at even larger disorders, with the conductance at the charge neutrality point of $E_{\rm F}/t=0$ being always larger than that at any other Fermi-level. However, in the spin-flip case, when the disorder strength exceeds a critical value, the quantized conductance at $E_{\rm F}/t=0$ abruptly vanishes, while the conductances at other energies remain finite at even larger disorders. We then show that this anomalous transport phenomenon in a mesoscopic system can be attributed to the exchange of Berry curvatures carried respectively by the conduction and valence bands in the corresponding bulk system. Based on the scaling theory of localization length, we provide a phase diagram to show the insulator-metal and metal-insulator phase transitions. At the end, a phenomalogical picture is given to understand the physical origin of the phase transitions from the spin-texture evolutions.

\textit{Anomalous Edge-State Transport.---} We start from a prototypical system with single massive Dirac fermion. Its corresponding tight-binding Hamiltonian in square lattices can be written as~\cite{QAHE-QiXL}:
\begin{eqnarray}
H=&-&\frac{v_{\rm F}}{2}\sum_i (c^\dag_i \sigma_{x} c_{i+\hat{x}} + c^\dag_i \sigma_y c_{i+\hat{y}} +h.c.) \nonumber
\\
&+&\frac{1}{2}\sum_{\langle ij \rangle} c^\dag_i \sigma_z c_{j} +\lambda \sum_{\langle i \rangle} c^\dag_i \sigma_z c_{i},
\end{eqnarray}
where $v_{\rm F}$ and $\lambda$ are Fermi velocity and the mass amplitude, which can be properly selected to produce a topologically trivial or nontrivial insulator. $t$ measures the nearest-neighbor hopping energy. In our consideration, we choose $v_{\rm F}=t$ and $\lambda=1.2t$ to realize the QAHE, giving rise to a quantized Hall conductance of $\sigma_{xy}=-e^2/h$. The applied Anderson disorders are included as $H_{\rm D}={\sum_i} w^0_i {c^\dag_i} c_{i} +w^x_i {c^\dag_i} \sigma_x c_{i}+w^y_i {c^\dag_i} {\sigma_y} {c_i}$, where the first term is the on-site nonmagnetic disorder, and the last two terms describe spin-flip disorders. And $w_{0,x,y}$ are uniformly distributed in an interval of [-$W$/2, $W$/2], with $W$ characterizing the disorder strength.

\begin{figure}
  \includegraphics[width=8cm,angle=0]{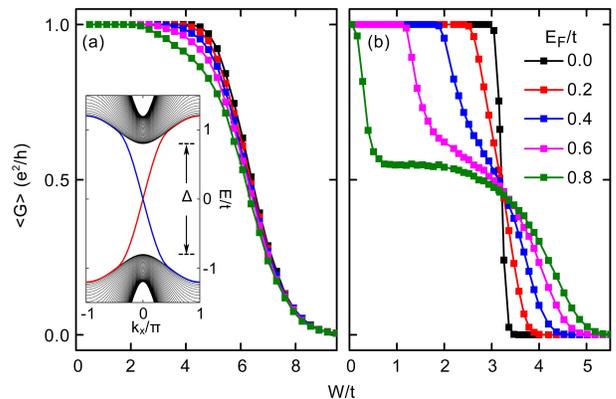}
  \caption{(color online) Averaged conductance $\langle G \rangle$ as a function of $W$ for different Fermi energies inside the bulk gap. The system width is set to be $N=80a$. (a) and (b): For nonmagnetic and spin-flip disorders, respectively. Over 3000 samples are collected for each point. Inset: Band structure of a nanoribbon displaying the gapless chiral edge states inside the bulk gap $\Delta$.} \label{figure1}
\end{figure}

Inset of Fig.~\ref{figure1}a plots the band structure of a nanoribbon of the QAHE system, where the ribbon width is set to be $N=80a$ ($a$ is the lattice constant). The gapless edge modes appear inside the bulk band gap of $E_{\rm F} \in[-0.8, 0.8]$ and exhibit chiral propagating characteristic~\cite{QHE3,QHE4}, with the red and blue respectively indicating the two counter-propagating edge modes along opposite boundaries. To explore the disorder effects on the QAHE, we use a two-terminal mesoscopic setup to study the averaged conductance $\langle G \rangle$ as a function of $W$. The disorders are only added in the central $N \times N$ scattering region connecting with left and right metallic semi-infinite terminals. Using the Landauer-B\"{u}ttiker formula~\cite{Datta}, the conductance $G$ can be evaluated as:
\begin{eqnarray}
  G=\frac{e^2}{h} {\rm{Tr}}[\Gamma_{\rm L}G^{r}\Gamma_{\rm R}G^{a}],
\end{eqnarray}
where $G^{r,a}$ are respectively the retarded and advanced Green's functions of the disordered region, and $\Gamma_{\rm L,R}$ are the line-width functions coupling left and right terminals with the central disordered region.

Since the conduction and valence bands are symmetric about $E_{\rm F}/t=0.0$, we choose five representative Fermi energies in our calculation, \textit{i.e.}, $E_{\rm F}/t$=0.0, 0.2, 0.4, 0.6, and 0.8. Figure~\ref{figure1}a displays the averaged conductance as a function of $W$ in the nonmagnetic case. At weak disorders, the conductance keeps quantized value at $\langle G \rangle=1.0 e^2/h$. When $W>2$, all conductances begin to gradually decrease with the increase of $W$. In particular, the conductance at $E_{\rm F}/t=0.0$ is always larger than those at other energies, which is rather reasonable because it requires much more energy to be scattered into the bulk. However, in the spin-flip case, the situation changes completely, with anomalous transport phenomena being observed~(See Fig.~\ref{figure1}b). For example, at $E_{\rm F}/t=0.0$ the conductance keeps quantized until the disorder reaches a critical strength of $W_{\rm C}/t\approx 3.2$, and abruptly vanishes when $W>W_{\rm C}$. While for other energies, the quantization of the conductance can be destroyed by weak disorders (\textit{i.e.}, the farther away from $E_{\rm F}/t=0.0$, the easier to be destroyed). However, the quantization becomes finite in a wide range of disorder strength. In contrast to the nonmagnetic disorder case, at the strong disorders $W>W_{\rm C}$, the closer of the Fermi energy to the charge neutrality point $E_{\rm F}/t=0.0$, the easier of the quantization of conductance being destroyed. To uncover the fundamental physics that results in these anomalous transport properties, we analyze the Berry curvature density in the corresponding bulk system that reflects the nature of the anomalous Hall effect~\cite{NiuRMP}, and employ the finite-size scaling theory~\cite{ScalingTheory} to determine the disorder-induced phase transitions in non-interacting electronic systems.

\begin{figure}
  \includegraphics[width=8cm,angle=0]{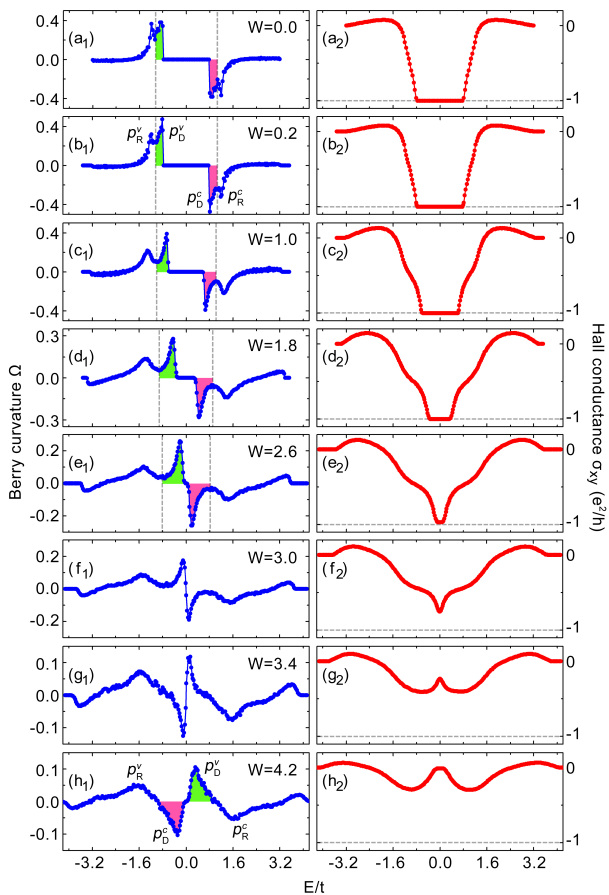}
  \caption{(color online) (a$_1$)-(h$_1$): Evolution of averaged Berry curvature density $\Omega$ as a function of energy $E$ for different disorder strengths $W$. The green and pink areas highlight the exchange process of Berry curvatures carried respectively by conduction and valence bands. (a$_2$)-(h$_2$): Corresponding averaged Hall conductance $\sigma_{xy}$ as a function of energy $E$. The supercell is set to be $50a\times 50a$. Over 30 samples are collected for each point.} \label{figure2}
\end{figure}

\textit{Berry Curvature Analysis.---} Instead of utilizing the Kubo formula for a clean system~\cite{KuboFormula}, here we adopt the generalized Berry curvature and Kubo formula for the disordered system, which can be respectively expressed as~\cite{Yang}:
\begin{eqnarray}
\Omega_{\alpha}&=&- \sum_{\beta \neq \alpha} \frac{2 {\rm {Im}}\langle \alpha |v_x|\beta\rangle \langle \beta|v_y| \alpha\rangle}{ (\omega_\beta - \omega_\alpha)^2}, \\
\sigma_{xy}&=&-\frac{e^2}{h}\int  {\rm d} \varepsilon \langle \Omega(\varepsilon) \rangle  f(\varepsilon)
\end{eqnarray}
where $|\alpha\rangle$ indicates the eigenenergy of $|\hbar \omega_{\alpha}\rangle$ in the disordered system, and $\Omega(\varepsilon)=\frac{1}{A} {\rm Tr}[ {\hat{\Omega}} \delta (\varepsilon-\hat{H})]$ describes the Berry curvature density in the energy spectrum with ${\hat{\Omega}}$ being the Berry curvature operator${\hat{\Omega}}=\sum_{\alpha} \Omega_{\alpha} |\alpha\rangle \langle\alpha|$, and  $A$ is the area of the two-dimensional system.

Figure~\ref{figure2} displays the averaged Berry curvature density and Hall conductance as functions of the energy $E/t$ for different spin-flip disorder strengths $W/t$=0.0, 0.2, 1.0, 1.8, 2.6, 3.0, 3.4, and 4.2. One can observe that even in the presence of strong disorders the averaged Berry curvature density and Hall conductance still satisfy the relations of  $\Omega(-E)=-\Omega(E)$ and $\sigma_{xy}(-E)=\sigma_{xy}(E)$,  in consistent with the averaged symmetric band structure in our system. At $W/t=0.0$, one can see that there are two Berry curvature peaks in either valence or conduction bands as shown Fig.~\ref{figure2}a$_1$, which can be generally identified as the contributions from both the massive Dirac bands (labelled as $p^{v,c}_{\rm D}$) and the remaining bands (labelled as $p^{v,c}_{\rm R}$), where $v/c$ denotes valence/conduction bands and $D/R$ denotes the Dirac/remaining bands. It is known that one continuum massive Dirac model of $h_0= \upsilon_F (\sigma_x k_x + \sigma_y k_y) +\frac{\lambda}{2} \sigma_z$ contributes to a half-quantized Hall conductance, \textit{i.e.}, $\sigma_{xy}=-\frac{1}{2} {\rm sgn} (\lambda)e^2/h$ for the parameters considered in our system, which is impossible in a non-interacting electronic system. Therefore, the contribution from the remaining bands must contribute another half Hall conductance as that from the massive Dirac bands in our considerations~\cite{QSHESCZ}. For energies inside the gap, $\Omega(E)=0$, and $\sigma_{xy}=-e^2/h$. When a weak disorder is applied, the bulk gap is nearly unaffected, but the peak $p^{v,c}_{\rm D}$ from the Dirac bands becomes rather singular (See Fig.~\ref{figure2}a$_2$), agreeing with the previous finding reported in Ref.~[\onlinecite{Yang}]. At moderate disorder strengths as shown in Figs.~\ref{figure2}b-~\ref{figure2}e, one can see that: 1) The peaks $p^{v,c}_{\rm D}$ become slightly broadened and move towards each other between conduction and valence bands in an attractive manner, shrinking the bulk gap; 2) The green and pink colored areas covered by the peaks $p^{v,c}_{\rm D}$ are approximately constants, corresponding to $\sigma_{xy}\approx \pm 0.5{e^2}/{h}$; and 3) the peaks $p^{v,c}_{\rm R}$ also broaden but move farther away from each other in a repulsive manner. Note that, before the bulk gap closing, the quantized Hall conductance at $E_{\rm F}=0$ is the most robust against disorders.
\begin{figure}
  \includegraphics[width=8cm,angle=0]{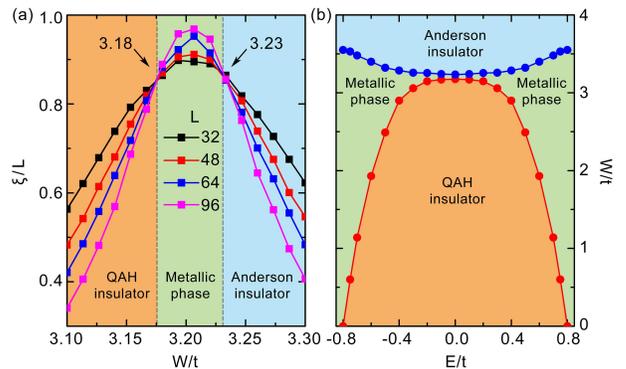}
  \caption{(color online) (a) Normalized localization length $\xi/L$ as a function of the disorder strength $W$ at $E_{\rm F}=0.0$ calculated on quasi-one dimensional bars, with a length of $2\times 10^6$ and different widths of $L=32$, 48, 64, and 96. $W_{\rm c1}=3.18$ and $W_{\rm c2}=3.23$ are two critical points. (b) Phase diagram of in ($E$, $W$) plane.} \label{Fig3}
\end{figure}

When the disorder strength approaches certain critical values, \textit{i.e.}, $W/t \approx 3.0$, the Berry curvatures from both valence and conduction bands become overlapped, closing the bulk band gap as displayed in Fig.~\ref{figure2}f. Surprisingly, we find that the Hall conductance at $E/t=0$ suddenly increases, faster than those at other energies. At even stronger disorders, \textit{e.g.}, $W/t=4.2$, one can see that the two peaks $p^{v,c}_{\rm D}$ from the conduction and valence bands make an exchange, accompanying with a bulk band gap reopening that is also confirmed from the band structure calculation~\cite{SeeSI}. When the bulk gap is reopened, the Hall conductance becomes exactly zero, \textit{i.e.}, $\sigma_{xy}=0$. Based on our integration, the Berry curvatures in the green/pink-colored regions contribute to a Hall conductance of about $-/+0.5e^2/h$. Therefore, we conclude that it is the exchange of the Berry curvatures carried respectively by the massive Dirac bands from the conduction and valence bands that leads to the compensation of the Berry curvatures carried by the conventional bands, which is a necessary condition to eliminate the Hall conductance (i.e. resulting in a \textit{net} Berry curvature integration below the Fermi level) and finally lead to the Anderson localization. We believe that such a feature of the Berry curvature exchange should be closely related to the anomalous transport behaviour shown in Fig.~\ref{figure1}b. In order to provide a convincing picture on the phase transitions from the QAHE to the Anderson insulator, it is a rewarding way to employ the finite-size scaling approach to determine the phase boundaries.
\begin{figure}
  \includegraphics[width=8cm,angle=0]{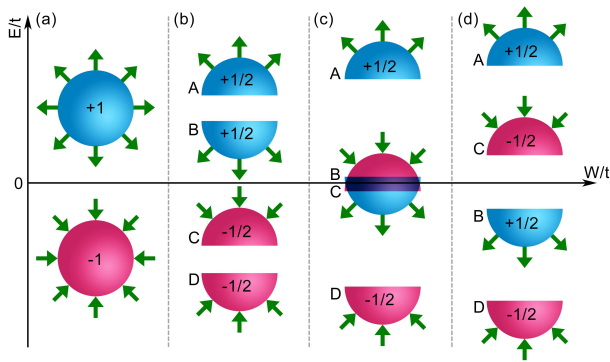}
  \caption{(color online) Schematic of the evolution of topological charges carried by the spin-textures (\textit{i.e.}, Skyrmions and Merons) of the valence and conduction bands. (a) In the absence of disorders, valence/conduction bands carry a Skyrmion with a topological charge of $Q_{\rm v}=-1$/$Q_{\rm c}=+1$. (b) At weak disorders, the states near the equator of Skyrmions scatter strongly to divide Skyrmions into Merons that carry half-integer topological charges and are labelled as ``A", ``B", ``C", and ``D". (c) By further increasing disorder strength, Merons A and D respectively move upwards and downwards, while Merons B and C move towards each other. (d) At even stronger disorders, Merons B and C make an exchange to cancel out the topological charges in the new valence (conduction) bands by combing Merons A and C (B and D).} \label{Fig4}
\end{figure}

\textit{Phase Transitions.---} Based on the well-established transfer-matrix method~\cite{LL1,LL2}, we numerically calculate the localization length $\xi$ on a quasi-one dimensional bar of essentially infinite length ($2 \times 10^6$) and finite width $L$. The periodical condition is applied to eliminate the possible edge-state transport. As a concrete example, in Fig.~\ref{Fig3}a we plot the normalized localization length $\xi/L$ as a function of $W/t$ at a fixed Fermi energy $E_{\rm F}=0.0$ for different widths $L$=32, 48, 64, and 96. One can find that there are two fixed points, $W_{\rm c1}=3.18$ and $W_{\rm c2}=3.23$. At $W< W_{\rm c1}$, $\xi/L$ decreases with the increase of $L$, indicating that $\xi/L$ will converge to zero when $L\rightarrow \infty$, signaling a localized insulating phase, \textit{i.e.}, the QAHE insulating phase. At $W\in  [ W_{\rm c1}, W_{\rm c2}]$, $\xi/L$ increases with the increase of $L$, indicating that $\xi/L$ will diverge when $L\rightarrow \infty$, signaling a delocalized metallic phase. At $W>W_{\rm c2}$, $\xi/L$ behaves similar as that in the weak disorder case, meaning that it enters a localized insulating phase (Anderson insulating phase). Therefore, the fixed points $W_{\rm c1}$ and $W_{\rm c2}$ are two critical disorder strengths for the insulator-metal and metal-insulator phase transitions, respectively.

After obtaining the two critical disorder strengths $W_{\rm c1}$ and $W_{\rm c2}$ for other representing Fermi energies inside the bulk gap, a phase diagram in the ($E$, $W$) plane can be determined (see Fig.~\ref{Fig3}b), which is also confirmed from the conductance calculation~\cite{SeeSI}. At weak disorders, the QAHE phase is robust against disorders, and the charge neutrality point $E_{\rm F}/t=0.0$ is most robust. However, at even stronger disorders, it is the charge neutrality point that first enters the Anderson insulating phase from a delocalized metallic phase. This anomalous feature of broadening the metallic phase associated with the Fermi-level shift from $E_{\rm F}/t=0.0$  is exactly the fundamental physical origin of the anomalous findings in the above two-terminal conductance calculation (See Fig.~\ref{figure1}b) and Hall conductance calculation in a finite-sized supercell (See Figs.~\ref{figure2}f$_2$ and \ref{figure2}g$_2$).

\textit{A Phenomenological Picture.---} We now provide a phenomenological picture to better explain the above observed anomalous transport findings and unusual phase diagram. From the topological point of view, the topological order of the quantized Hall conductance can also be described by using the concept of topological charge, which is dependent on the spin-textures~\cite{Skyrmion}. In our consideration, the quantized Hall conductance of $\sigma_{xy}=-e^2/h$ is analogous to a Skyrmion, where the valence/conduction bands carry a topological charge of $Q_{\rm v/c}=-/+1$, which can be reflected from the spin-textures as schematically displayed in Fig.~\ref{Fig4}a: spins pointing outwards for $Q_{\rm c}=+1$ (\textit{i.e.}, spins point upwards at the north pole, downwards at the south pole, and in-plane at the equator), but spins pointing inwards for $Q_{\rm v}=-1$ (\textit{i.e.}, spins point downwards at the north pole, upwards at the south pole, and in-plane at the equator). When the spin-flip disorders are applied, the electronic states near the equator scatter strongly due to the coexistence of different spins, while those at the south and north pole are nearly inactive because of the absence of allowed states with opposite spin. This results in the separation of Skyrmions into Merons (labelled as ``A", ``B", ``C", and ``D") with $\pm 1/2$ topological charges as labelled in Fig.~\ref{Fig4}b. With the increase of disorder strength, Merons ``B" and ``C" move towards each other in an attractive manner, while Merons ``A" and ``D" move in a repulsive manner as displayed in Fig.~\ref{Fig4}c. Note that, at certain disorder strength, Merons ``B" and ``C" become spatially overlapped, but no scattering occurs because it is not allowed to scatter between same spins, which explains the formation of the Berry-curvature mediated metallic phase in Fig.~\ref{Fig3}. At even larger disorder strength, Merons ``B" and ``C" move respectively into the valence and conduction bands. At this point, the topological charges carried by the occupied valence bands becomes zero, leading to the vanishing of Hall conductance and the occurrence of Anderson localization, which can clearly explain the sudden vanishing of the two-terminal conductance and why the charge neutrality point enters the Anderson insulator first.

\textit{Conclusions.---} In summary, we theoretically study the disorder effect of the QAHE system in the presence of spin-flip disorders. We show that the system first transitions into a metallic phase from the QAHE insulating phase at moderate disorder strengths, and then further transitions into the Anderson insulating phase. Counterintuitively, we find that it is the charge neutrality point (\textit{i.e., $E_{\rm F}/t=0$}), at which the quantized Hall conductance is most robust against weak disorders but the corresponding metallic phase is easiest to be completely localized into the Anderson insulating phase. By analyzing the Berry curvature evolution, we find that the resulting anomalous electronic transport and Anderson localization at the charge neutrality point originate from the \textit{interchange} of Berry curvatures carried respectively by the valence and conduction bands. The finite-size localization length scaling approach is used to determine the phase boundaries separating the three phases: QAHE insulating phase, Berry-curvature mediated metallic phase, and the Anderson insulating phase. At the end, a phenomenological picture from the topological charges is given to explain the anomalous electronic transport and Anderson localization at the charge neutrality point.

\textit{Acknowledgements.---} We are grateful to D.-N. Sheng, J.-R. Shi, and A. H. MacDonald for valuable discussions. This work was financially supported by the China Government Youth 1000-Plan Talent Program, Fundamental Research Funds for the Central Universities (WK3510000001 and WK2030020027), NNSFC (Grant Nos. 11474265, 11374219, and 91121004), and NBRPC (2013CB921900 and 2012CB921300). S.A.Y. thanks the financial support by Singapore University of Technology and Design (SUTD-SRG-EPD2013062 and SUTD-T1-2015004). J.W. thanks the financial support by the University Grant Council (AoE/P-04/08) of the Government of HKSAR. Q.N. also acknowledges financial support from DOE (DE-FG03-02ER45958, Division of Materials Science and Engineering) and the Welch Foundation (F-1255). The Supercomputing Center of USTC is gratefully acknowledged for the high-performance computing assistance.

\end{document}